\documentclass[pra, aps, twocolumn, preprintnumbers,showpacs, floatfix,superscriptaddress]{revtex4}
\usepackage{hyperref}
\usepackage{latexsym}
\usepackage{mathrsfs}
\usepackage{amsmath}
\usepackage{amssymb}
\usepackage{amsfonts}
\usepackage{ifthen}
\usepackage{graphicx}% Include figure files
\usepackage{dcolumn}% Align table columns on decimal point
\usepackage{bm}% bold math
\usepackage{color}
\begin{document}

\title{Semi-quantum secret sharing using entangled states}% Force line breaks with \\

\author{Qin Li}

\affiliation{Department of Computer Science, Sun Yat-sen University, Guangzhou 510275, China}%Lines break automatically or can be forced with \\

\author{W. H. Chan}
\affiliation{Department of Mathematics, Hong Kong Baptist
University, Kowloon, Hong Kong, China}

\author{Dong-Yang Long}
\affiliation{Department of Computer Science, Sun Yat-sen University,
Guangzhou 510275, China}

\date{\today}% It is always \today, today,

\begin{abstract}
Secret sharing is a procedure for sharing a secret among a number of
participants such that only the qualified subsets of participants
have the ability to reconstruct the secret. Even in the presence of
eavesdropping, secret sharing can be achieved when all the members
are quantum. So what happens if not all the members are quantum? In
this paper we propose two semi-quantum secret sharing protocols using maximally entangled GHZ-type states in which quantum Alice shares a secret with two
classical parties, Bob and Charlie, in a way that both parties are
sufficient to obtain the secret, but one of them cannot. The
presented protocols are also showed to be secure against
eavesdropping.
\end{abstract}
%\keywords{Suggested keywords}%Use show keys class option if keyword
\maketitle

\section{Introduction}
Suppose a service provider Alice wants to distribute some secret
information among clients, Bob and Charlie, such that Bob and
Charlie can obtain the secret information through the cooperation,
while one of them cannot. Classical secret sharing has been proposed
as a solution \cite{bla,sha,sg}. A simple example is that Alice
prepares a binary bit string related to her secret message and
generates a random string of the same length, applies bitwise XOR
operations on such two strings, and then sends the resulting string
to Bob and a copy of the random string to Charlie. Obviously, Bob
and Charlie acting together can access to Alice's message, but one
of them can obtain nothing about it.

Unfortunately, classical secret sharing cannot address the problem
of eavesdropping if it is not used in conjunction with other
techniques such as encryption. If an eavesdropper Eve (including one
malicious participant of the Bob-Charlie pair) can control the
communication channel and obtain both of Alice's transmissions, then
Alice's message becomes transparent for her. Fortunately, quantum
secret sharing can achieve secret sharing and eavesdropping
detection simultaneously. Hillery \emph{et al}. showed how to
implement a secret sharing scheme using three-particle entangled
Greenberger-Horne-Zeilinger (GHZ) states \cite{ghz} in the presence
of an eavesdropper \cite{hbb}. Karlsson \emph{et al}. presented a
secret sharing scheme based on two-particle quantum entanglement
such that only two members implementing together are able to obtain
the information \cite{kki}. Gottesman showed that the size of each
important share sometimes can be made half of the size of the secret
if quantum states are used to share a classical secret \cite{dg}.
The secret sharing protocol among $n$ parties based on entanglement
swapping of $d$-level cat states and Bell states was introduced by
Karimipour \emph{et al}. \cite{kbb}. Guo \emph{et al}. proposed a
secret sharing scheme utilizing product states instead of entangled
states and thus the efficiency is improved to approach 100\%
\cite{gg}. Xiao \emph{et al.} generalized the scheme in Ref.
\cite{hbb} into any number of participants and gave two efficient
quantum secret sharing schemes with the efficiency asymptotically
100\% \cite{xldp}. Zhang \emph{et al}. considered a multiparty
quantum secret sharing protocol of the classical secret based on
entanglement swapping of Bell states \cite{zm}. There are also many
quantum secret sharing protocols considering sharing quantum
information \cite{hbb,kki,dg,cgl,sb,nmi,lzp,lsb,zlm,sbz,mp,ms}.
Especially, Markham \emph{et al}. developed a unified approach to
secret sharing of both classical and quantum secrets employing graph
states \cite{ms}.

However, previous quantum secret sharing protocols requires all the
parties to have quantum capabilities. So what happens if not all the
parties are quantum? Actually, the situation that not all the
participants can afford expensive quantum resources and quantum
operations is more common in various applications. It is well known
that semi-quantum key distribution in which one party Alice is
quantum and the other party Bob just owns classical capabilities is
possible \cite{bgkm,bkm1,bkm2}, so it is interesting to ask whether
semi-quantum secret sharing (the specific definition is given
afterwards) is possible. The answer is affirmative.

In this paper, we consider the secret sharing protocol in which
quantum Alice has to share a secret with classical Bob and classical
Charlie such that the collaboration of Bob and Charlie can
reconstruct the secret, while one of them cannot obtain anything
about the secret. We say Alice is quantum when she is allowed to
prepare arbitrary quantum states and perform any quantum operations.
We follow the descriptions about ``classical'' in Refs.
\cite{bkm1,bkm2,bgkm}. The computation basis
$\{|0\rangle,|1\rangle\}$ is called ``classical'' and is replaced
with the classical notations $\{0,1\}$. Bob and Charlie is classical
when they are restricted to performing four operations when they
access a segment of the quantum channel: (1) measuring the qubits in
the classical basis $\{0,1\}$; (2) reordering the qubits (via proper
delay measures); (3) preparing (fresh) qubits in the classical basis
$\{0,1\}$; (4) sending or returning the qubits without disturbance.
The protocol of this kind is termed as ``Semi-Quantum Secret Sharing
(SQSS)''. SQSS protocols can have two variants,
\emph{randomization-based} SQSS and \emph{measure-resend} SQSS, in
terms of the operations which classical participants are allowed to
implement. In a randomization-based SQSS protocol, classical
participants are limited to perform the operations (1), (2), and
(4), while in a measure-resend SQSS protocol, classical participants
are limited to perform the operations (1), (3), and (4). In
principle, a SQSS protocol is considered as secure if neither an
eavesdropper nor a malicious participant can obtain any information
about the secret. In the following section, we utilize maximally entangled states of the GHZ type to
construct a randomization-based SQSS protocol and a measure-resend
SQSS protocol based on the semi-quantum key distribution protocols
\cite{bkm1,bkm2,bgkm}, and we show that the proposed SQSS protocols
are secure against eavesdropping.

\section{Three-particle entangled states}
In order to construct semi-quantum secret sharing protocols, we
introduce a three-particle maximally entangled state in the following form
\begin{equation}
|\psi\rangle=\frac{1}{\sqrt
2}(|0\rangle\frac{|00\rangle+|11\rangle}{\sqrt
2}+|1\rangle\frac{|01\rangle+|10\rangle}{\sqrt 2}).\label{eq:one}
\end{equation} This state also can be rewritten as
\begin{eqnarray}
|\psi\rangle &=& \frac{1}{\sqrt
2}(|0\rangle\frac{|++\rangle+|--\rangle}{\sqrt
2}+|1\rangle\frac{|++\rangle-|--\rangle}{\sqrt 2}) \nonumber \\
&=&\frac{|0\rangle+|1\rangle}{\sqrt2}\frac{|++\rangle}{\sqrt2}+
\frac{|0\rangle-|1\rangle}{\sqrt2}\frac{|--\rangle}{\sqrt 2}\nonumber \\
&=&\frac{|+++\rangle+|---\rangle}{\sqrt2}.\label{eq:two}
\end{eqnarray} Obviously, by implementing Hadamard operation on each particle of the state $|\psi\rangle$ respectively, $|\psi\rangle$ is transformed into the standard GHZ state, $|GHZ\rangle=\frac{|000\rangle+|111\rangle}{\sqrt2}$. According to Ref. \cite{dvc},  if two three-particle entangled states can be mutually transformed by local unitary operations, they are equivalent. Hence, as an entangled state, $|\psi\rangle$ is equivalent to the standard GHZ state and belongs to the GHZ type.

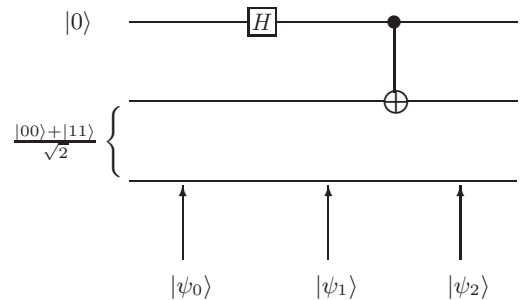
\begin{figure}
\begin{picture}(215,120)
\put(40,105){$|0\rangle$}\put(65,107.5){\line(1,0){44}}
\put(110,102.5){\framebox(10,10){\emph{H}}}\put(120,107.5){\line(1,0){44}}
\put(165,107.5){\circle*{5}}\put(167.5,107.5){\line(1,0){44}}
\put(20,60){$\frac{|00\rangle+|11\rangle}{\sqrt 2}$}
\put(55,60){\Bigg\{}\put(65,77.5){\line(1,0){96}}\put(160.5,75){$\bigoplus$}
\put(165,81){\line(0,1){27}}\put(169.5,77.5){\line(1,0){42}}
\put(65,47.5){\line(1,0){146}} \put(85,17.5){\vector(0,1){28}}
\put(140,17.5){\vector(0,1){28}}\put(190,17.5){\vector(0,1){28}}
\put(80,5){$|\psi_0\rangle$}\put(135,5){$|\psi_1\rangle$}
\put(185,5){$|\psi_2\rangle$}
\end{picture}
\caption{Quantum circuit for generating $|\psi\rangle$}\label{fig:1}
\end{figure}

The GHZ-type state $|\psi\rangle$ is not only theoretically existent
but also practically feasible. It can be obtained from the standard GHZ state, and also can be generated in the following way. To gain $|\psi\rangle$, we may
begin with preparing the state $|0\rangle$ and the Bell state
$\frac{|00\rangle+|11\rangle}{\sqrt 2}$, and then apply the Hadamard
gate to the first qubit, and finally apply the controlled-NOT gate
to the first two qubits. The specific steps are illustrated by the
quantum circuit showed in Figure \ref{fig:1}. Let us follow the
states in the circuit to see clearly the process of generating
$|\psi\rangle$. The input state of circuit is
\begin{equation}
|\psi_0\rangle=|0\rangle\otimes \frac{|00\rangle+|11\rangle}{\sqrt
2}.
\end{equation}
After sending the first qubit through the Hadamard gate, we have
\begin{eqnarray}
|\psi_1\rangle &=& \frac{|0\rangle+|1\rangle}{\sqrt 2}\otimes
\frac{|00\rangle+|11\rangle}{\sqrt 2} \nonumber\\
&=& \frac{1}{\sqrt 2}(|0\rangle \frac{|00\rangle+|11\rangle}{\sqrt
2}+|1\rangle \frac{|00\rangle+|11\rangle}{\sqrt 2}).
\end{eqnarray}
Then we send the first two qubits through the controlled-NOT gate to
obtain
\begin{eqnarray}
|\psi_2\rangle &=& \frac{1}{\sqrt 2}(|0\rangle
\frac{|00\rangle+|11\rangle}{\sqrt 2}+|1\rangle
\frac{|10\rangle+|01\rangle}{\sqrt 2}) \nonumber \\
&=& |\psi\rangle.
\end{eqnarray}

\section{Randomization-based SQSS protocol}
In this section, we propose a randomization-based SQSS protocol in
which quantum Alice and the other two classical parties, Bob and
Charlie, share a secret string such that Bob and Charlie can recover
the secret string only when they work together. Quantum Alice has
the ability to prepare the maximally entangled GHZ-type state
$|\psi\rangle$ and perform some quantum operations such as Bell
measurements and three-particle measurements. Classical parties, Bob
and Charlie, are restricted to implementing three operations: (1)
measuring the qubits in the classical basis $\{0,1\}$; (2)
reordering the qubits (via proper delay measures); (3) sending or
returning the qubits without disturbance. All the participants can
access a quantum channel and an authenticated public channel that is
susceptible to eavesdropping. The detailed steps are given in the
following.

1. Alice creates a sufficiently long string of three-particle
entangled states in the form of Eq. (\ref{eq:one}) (Suppose $N$
triplet states are in the string and theses states are indexed from
$1$ to $N$). After that, Alice sends the second and the third
particle of each entangled state to Bob and Charlie, and keeps
the remaining for herself.

2. Upon receiving each qubit, Bob randomly determines, either to
measure the qubit using the classical basis $\{0,1\}$ (we refer to
this action as ``SHARE''), or to reflect it back to Alice (we refer
to this action as ``CHECK''). Particularly, Bob reflects the qubits
in a new order such that nobody else could distinguish which qubits
are returned. Each measurement outcome is interpreted as a binary
$0$ or $1$. Similarly, Charlie also randomly decides either to
measure the qubits or to reflect the qubits in another order.

3. Alice temporarily restores the qubits reflected by Bob and Charlie in quantum registers according to their incoming sequences, and announces
that she has received their reflected particles in a public channel.

4. Bob and Charlie declare which qubits were reflected by them and
the order in which their qubits were returned, respectively.

5. For her own qubit in each position, Alice performs one of the
four actions according to Bob's and Charlie's actions on the
corresponding qubits, as illustrated in Table \ref{tab:1}.
\begin{table}
\caption{\label{tab:1}Participants' actions on the qubits in each
position}
\begin{ruledtabular}
\begin{tabular}{c|c|c|c}
 Case & Bob & Charlie & Alice\\
  \hline
  (1) & SHARE & SHARE & ACTION 1 \footnote{Measuring her qubit in the classical basis $\{0,1\}$} \\
  (2) & SHARE & CHECK & ACTION 2 \footnote{Combining her qubit with Charlie's reflected qubit and performing a Bell measurement} \\
  (3) & CHECK & SHARE & ACTION 3 \footnote{Combining her qubit with Bob's reflected qubit and performing a Bell measurement} \\
  (4) & CHECK & CHECK & ACTION 4 \footnote{Combining her qubit with the two reflected qubits and performing an appropriate three-particle measurement} \\
\end{tabular}
\end{ruledtabular}
\end{table}

It is supposed that there are four cases appearing in the same
probability: (1) both Bob and Charlie choose to SHARE, then Alice
can implement ACTION 1 to obtain a bit (we name this bit as SHARE
bit) that can be retrieved if Bob and Charlie use the XOR operation
on their measurement outcomes; (2) Bob chooses to SHARE and Charlie
chooses to CHECK, then Alice can perform ACTION 2 to check whether
Bob's measurement outcome is right and the resulting two-particle
state is the correct Bell state; (3) Bob chooses to CHECK and
Charlie chooses to SHARE, then Alice can utilize ACTION 3 to check
if Charlie's measurement result is right and the resulting
two-particle state is the correct Bell state; and (4) both Bob and
Charlie choose to CHECK, then Alice can check whether the original
three-particle entangled states in the form of Eq. (\ref{eq:one}) is
changed by carrying out ACTION 4.

For instance, let Bob randomly measure the qubits in $N/2$ positions
(SHARE) and reflect the qubits in the other $N/2$ positions in a new
order $l_B=l_1l_2\cdots l_{N/2}$ (CHECK), and Charlie performs the
similar operations as Bob does and reflects the qubits in another
order $m_C=m_1m_2\cdots m_{N/2}$. Suppose $N=8$, $l_B=4731$ and
$m_C=6427$. Then the lists of the qubits measured by Bob and Charlie
are indexed by their complements $\bar{l}_B=2568$ and
$\bar{m}_C=1358$, respectively. Hence Alice performs ACTION 1 in the
positions 5 and 8 and interprets the measurement outcomes as
classical bits $0$ or $1$, and performs ACTION 4 in the positions 4
and 7. Alice also implements ACTION 2 in the positions 2 and 6 and
ACTION 3 in the positions 1 and 3.

6. Alice checks the error rate in cases (2), (3), and (4) given in Table \ref{tab:1}. If the
error rate in any case is higher than some predefined threshold
value, the protocol aborts.

7. Alice requires Bob and Charlie to reveal a random subset (assume
the size of the subset is about $N/8$) of the bits which are used to
generate Alice's SHARE bits. Actually this process is used to check the error rate in case (1). If the values of Bob's and Charlie's
bits are the same (or opposite), then Alice's bit should be $0$ (or
1) according to the Eq. (\ref{eq:one}). From step 5, we know
that approximately $N/4$ positions are selected by both Bob and
Charlie to SHARE. If the error rate on SHARE bits is not
significant, the remaining $N/8$ SHARE bits of Alice forms the final
secret string which can be recovered only when Bob and Charlie work
together.

We show the above randomization-based SQSS protocol is secure
against eavesdropping in two situations. The first is that one
dishonest classical party Bob (or Charlie) attempts to find Alice's
secret without cooperating with the other party in the recovery
stage. The second is that a fourth eavesdropper Eve who has quantum
capabilities is involved and aims to find Alice's secret without
being detected.

We first suppose the dishonest classical party Bob can access both
of Alice's transmissions. In some of the positions, Bob may measure
both qubits using the classical basis $\{0,1\}$ and resend one of
them in the state he found to Charlie. In terms of the Eq.
(\ref{eq:one}), if both of the measurement outcomes are the same (or
opposite), he learns that Alice's bit must be $0$ (or $1$). In the
other positions, Bob may behave like a honest party and do nothing
on Charlie's qubits. However, this cheating strategy can hardly
succeed since Bob does not know Charlie's choices. If Bob measures
Charlie's qubit in the position where Charlie chooses to CHECK, he
suffers a problem. According to the state $|\psi\rangle$ in Eq.
(\ref{eq:one}), if Bob just measures his own qubit, then the
two-particle state resulting from combining Alice's qubit and
Charlie's reflected qubit should be the Bell state, while if Bob
measures both qubits of him and Charlie, then the two-particle state
resulting from combining Alice's qubit and Charlie's reflected qubit
will be the product state, and thus Alice will find this abnormity
with probability $1/2$ using a Bell measurement. But if Bob measures
Charlie's qubit in the position where Charlie chooses to SHARE, his
cheating will not be found. In each position, Charlie has a
probability of $1/2$ of making either choice, so the probability
that Bob escapes detection is $1/2\times1/2+1/2=3/4$. Assume that
Bob has to measure both qubits in $l(l\leq N/4)$ positions to obtain
the significant information of Alice's secret without the aid of
Charlie, then the probability that Bob goes undetected is $(3/4)^l$
which may be arbitrarily small by picking an appropriate $l$ and
$N$.

Now let us consider the second case in which a fourth party Eve who
has quantum capabilities is involved. Assume that Eve can obtain
both of Alice's transmissions and tries to obtain Alice's secret. If
Eve gets Bob's and Charlie's qubits of certain entangled states, she
may measure the two qubits in the Bell basis and then resend the
qubits in the states she found to Bob and Charlie, respectively. In
terms of Eq. (\ref{eq:one}), if the measurement outcome is
$\frac{|00\rangle+|11\rangle}{\sqrt 2}$, Eve learns that Alice's bit
should be $0$; otherwise she knows that Alice's bit should be $1$.
However, Eve's cheating is likely to be detected since she does not
know Bob's and Charlie's choices. If she measures the qubits in the
position where both Bob and Charlie choose to CHECK, then the
three-particle state resulting from combining Alice's qubit and the
other two reflected qubits will be the product of one single state
and a Bell state but not the same as the original state
$|\psi\rangle$ in the form of Eq. (\ref{eq:one}), and thus Alice can
discover this defraud with the probability $1/2$ by measuring it in
a three-particle orthogonal basis
$\{|\phi_0\rangle,|\phi_1\rangle,...,|\phi_7\rangle\}$, where
\begin{equation}
|\phi_0\rangle=\frac{1}{\sqrt
2}(|0\rangle\frac{|00\rangle+|11\rangle}{\sqrt
2}+|1\rangle\frac{|01\rangle+|10\rangle}{\sqrt 2}),\nonumber
\end{equation}
\begin{equation}
|\phi_1\rangle=\frac{1}{\sqrt
2}(|0\rangle\frac{|00\rangle+|11\rangle}{\sqrt
2}-|1\rangle\frac{|01\rangle+|10\rangle}{\sqrt 2}),\nonumber
\end{equation}
\begin{equation}
|\phi_2\rangle=\frac{1}{\sqrt
2}(|0\rangle\frac{|00\rangle-|11\rangle}{\sqrt
2}+|1\rangle\frac{|01\rangle-|10\rangle}{\sqrt 2}),\nonumber
\end{equation}
\begin{equation}
|\phi_3\rangle=\frac{1}{\sqrt
2}(|0\rangle\frac{|00\rangle-|11\rangle}{\sqrt
2}-|1\rangle\frac{|01\rangle-|10\rangle}{\sqrt 2}),\nonumber
\end{equation}
\begin{equation}
|\phi_4\rangle=\frac{1}{\sqrt
2}(|1\rangle\frac{|00\rangle+|11\rangle}{\sqrt
2}+|0\rangle\frac{|01\rangle+|10\rangle}{\sqrt 2}),\nonumber
\end{equation}
\begin{equation}
|\phi_5\rangle=\frac{1}{\sqrt
2}(|1\rangle\frac{|00\rangle+|11\rangle}{\sqrt
2}-|0\rangle\frac{|01\rangle+|10\rangle}{\sqrt 2}),\nonumber
\end{equation}
\begin{equation}
|\phi_6\rangle=\frac{1}{\sqrt
2}(|1\rangle\frac{|00\rangle-|11\rangle}{\sqrt
2}+|0\rangle\frac{|01\rangle-|10\rangle}{\sqrt 2}),\nonumber
\end{equation}
\begin{equation}
|\phi_7\rangle=\frac{1}{\sqrt
2}(|1\rangle\frac{|00\rangle-|11\rangle}{\sqrt
2}-|0\rangle\frac{|01\rangle-|10\rangle}{\sqrt 2}).
\end{equation}
Similarly, if Eve measures the qubits either in the position where
Bob chooses to SHARE and Charlie chooses to CHECK, or in the
position where Bob selects to CHECK and Charlie selects to SHARE,
she also can be detected with probability $1/2$ by implementing a
Bell measurement. But if Eve measures the qubits in the position
where both Bob and Charlie choose to SHARE, she cannot be detected.
In every position, as Bob and Charlie have a probability of $1/2$ of
choosing to SHARE or CHECK, the probability that Eve's cheating is
undetected is $1/4\times1/2\times3+1/4=5/8$. Suppose there are
$m(m\leq N/4)$ positions where Bob should measure the qubits in a
Bell basis to learn the considerable information of the secret, then
Bob's cheating goes undetected with probability $(5/8)^m$ which can
be small enough by choosing a suitable $m$ and $N$. In addition, Eve
also obtains nothing about Alice's secret information even if she
manages to entangle an ancilla with each qubit of Bob (or Charlie).
Suppose that in a certain position, Eve has entangled an ancilla
$|0\rangle$ with Bob's qubit, and both Bob and Charlie measure their
qubits, then the Alice-Eve system collapses to $|00\rangle$ or
$|10\rangle$, which leaks no information to Eve about Alice's qubit.

Particularly, notice that it is indispensable for Alice to announce
that she has received all the reflected particles in step 3. If Eve
can learn which qubits were reflected by Bob and Charlie and in
which order they were reflected before Alice receives the reflected
qubits, he can obtain the secret string of Alice without inducing
errors by using the similar way to attack the mock protocol in Refs.
\cite{bgkm,bkm1,bkm2}. For each incoming qubit of Bob, she entangles
an ancilla $|0\rangle$ with it and implements a controlled-NOT
operation on them (Bob's qubit as the control qubit and the ancilla
qubit as the target qubit). Then she holds all the qubis that Bob
reflected until Bob publishes which qubits were reflected and in
which order they were reflected. Next she rearranges the reflected
quits in the same order as Alice sent them to Bob and performs
another controlled-NOT operation on each returned qubit and the
corresponding ancilla. After that, she resends the resulting qubits
in the order that Bob declared to Alice. Finally, in the position
where Bob chose to SHARE, she measures her ancilla and learns Bob's
bit. For the qubits sent to Charlie, Eve does the similar operations
and learns Charlie's bits. In the position where both Bob and Charlie
chose to SHARE, Eve can obtain the SHARE bit by implementing XOR
operation on their bits according to Eq. (\ref{eq:one}). Moreover,
Eve goes undetected since she introduces no errors.

\section{Measure-resend SQSS protocol}
In the following, a measure-resend SQSS protocol is introduced.
Quantum Alice can prepare the three-particle GHZ-type state
$|\psi\rangle$ and perform some quantum operations and classical
parties, Bob and Charlie, are restricted to performing three
operations: (1) measuring the qubits in the classical basis
$\{0,1\}$; (2) preparing (fresh) qubits in the classical basis
$\{0,1\}$; (3) sending or returning the qubits without disturbance.
This protocols is quite similar to the randomization-based SQSS
protocol except that step 2 and step 4 are adapted to the different
restrictions of classical participants, so the modified steps are
given as follows:

2. When Bob (or Charlie) receives each qubit he randomly determines,
either to measure it in the classical basis $\{0,1\}$ and return it
in the same state he found (SHARE), or to reflect it directly
(CHECK).

4. Bob and Charlie declare the positions in which the qubits were
measured (or reflected).

The proposed measure-resend SQSS protocol is secure against
eavesdropping in a way similar to that in the randomization-based
SQSS protocol. A dishonest party Bob (or Charlie) should not find
Alice's secret without collaborating with the other party and a
fourth eavesdropper Eve who has quantum capabilities also should not
obtain Alice's secret without disturbance. Suppose Bob is dishonest
and he has controlled both of Alice's transmissions. In some of the
positions, Bob measures both particles and resends one of them to
Charlie. However, if Charlie does not measure the qubits in such
positions, the Alice-Charlie systems should collapse to the Bell
states but not product states, which might be discovered by Alice
through implementing Bell measurements. Likewise, assume that a
fourth party Eve who owns quantum capabilities has managed to obtain
both Bob's and Charlie's particles. In certain positions, Eve
measures the two qubits in the Bell basis and then resends the
qubits to Bob and Charlie, respectively. However, if either Bob or
Charlie does not measure their qubits in such positions, Eve's
cheating will be detected by Alice through performing appropriate
measurements. Besides, even if Eve manages to entangle an ancilla
with each qubit of Bob (or Charlie), she also obtains nothing about
Alice's secret since the ancilla is always left unchanged.

Note that it is also significant to demand Alice to publish that she
has received all the reflected qubits in step 3 of this protocol. If
this requirement is loss, Eve can cheat successfully. For instance,
Eve holds the reflected qubits from Bob and Charlie until they
announce the positions in which the qubits were measured and resent
(SHARE), or reflected directly (CHECK). Then Eve measures the qubits
that they measured and then resends them in the states she found,
and reflects the qubits that they reflected without disturbance. In
the position where both Bob and Charlie measured their qubits, if
Eve's measurements are the same, then she learns Alice's bit must be
$0$; otherwise she learns Alice's bit must be $1$. Furthermore, Eve
can escape detection since she does not introduce disturbance
anywhere.

\section{Conclusion and discussion}
We have introduce a maximally entangled GHZ-type state and shown that it
is not only theoretically existent but also practically feasible. Furthermore,
we has used such GHZ-type states to propose two
semi-quantum secret sharing protocols in which Alice has quantum
capabilities, while the other two parties, Bob and Charlie, are
limited to classical operations: measure qubits in the classical
basis $\{0,1\}$; send or reflect qubits without disturbance; reorder
some qubits or prepare fresh qubits after measurements and resend
them. The proposed protocols also have been showed to be secure against eavesdropping.
Since the proposed SQSS protocols do not require all the
participants owning quantum capabilities, the secret sharing can be
achieved at a lower cost. Therefore, the applicability of secret
sharing could be widen to the situation in which not all the
participants can afford expensive quantum resources and quantum
operations.

Nevertheless, we just consider the case that quantum Alice shares a
secret with two classical parties, Bob and Charlie. An interesting
question is: can a general SQSS protocol in which quantum Alice
shares a secret with several parties who may be quantum or classical
be achieved? Besides, note that no noise were assumed and so that
three participants, namely, Alice, Bob, and Charlie, can share
perfect entangled states if eavesdroppers introduce no errors. So
another interesting question is: can entangled states can be shared
of almost perfect fidelity if not all the parties are quantum when
noisy quantum channels are used?

\section*{ACKNOWLEDGMENTS}
We would like to appreciate L. Lin, J. Zhang, J. X. Li, Y. X. Long,
and B. H. Wang for helpful discussions and suggestions. This work
was sponsored by the National Natural Science Foundation of China
(Project No. 60573039), and the Faculty Research (Grant No.
FRG2/08-09/070) Hong Kong Baptist University.

\end{document}